\documentclass[12pt]{article}

\usepackage{times}
\usepackage{amsthm,amscd,amsxtra,amsfonts,amsmath,amssymb,multirow}
\usepackage{algorithm,algorithmic}
\usepackage{bbm}
\usepackage{graphicx}
\usepackage{subcaption}
\usepackage{hyperref}
\usepackage{epstopdf}
\usepackage[export]{adjustbox}
\usepackage{xcolor}

\theoremstyle{definition}

\theoremstyle{remark}

\theoremstyle{example}

\theoremstyle{conjecture}

\newenvironment{sciabstract}{%
\begin{quote} \bf}
{\end{quote}}

\title{Ollivier persistent Ricci curvature (OPRC) based molecular representation for drug design}

\author
{JunJie Wee,$^{1}$ and Kelin Xia$^{1\ast}$\\
\\
\normalsize{$^{1}$Division of Mathematical Sciences, School of Physical and Mathematical Sciences,} \\
 \normalsize{Nanyang Technological University, Singapore 637371}\\
}

\date{}

\begin{document}


\baselineskip24pt


\maketitle

\begin{sciabstract}
Efficient molecular featurization is one of the major issues for machine learning models in drug design. Here we propose persistent Ricci curvature (PRC), in particular Ollivier persistent Ricci curvature (OPRC), for the molecular featurization and feature engineering, for the first time. Filtration process proposed in persistent homology is employed to generate a series of nested molecular graphs. Persistence and variation of Ollivier Ricci curvatures on these nested graphs are defined as Ollivier persistent Ricci curvature. Moreover, persistent attributes, which are statistical and combinatorial properties of OPRCs during the filtration process, are used as molecular descriptors, and further combined with machine learning models, in particular, gradient boosting tree (GBT). Our OPRC-GBT model is used in the prediction of protein-ligand binding affinity, which is one of key steps in drug design. Based on three most-commonly used datasets from the well-established protein-ligand binding databank, i.e., PDBbind, we intensively test our model and compare with existing models. It has been found that our model are better than all machine learning models with traditional molecular descriptors.
\end{sciabstract}

\section*{Introduction}

Drug design and discovery is one of the most important and challenging tasks in biological sciences. However, traditional drug design approaches with Edisonian trial-and-error experimental cycles are expensive and highly inefficient. In fact, nine out of ten candidate therapies have failed somewhere between phase I trials and regulatory approval. Since 1980, computer-aided drug design (CADD) \cite{macalino2015role}, which is to use computer-based models, methods, and algorithms from data-mining, pattern recognition, statistic learning, as well as biophysics and biochemistry, has been introduced to systematically analyze drug-related data. CADD has been involved in almost the whole pipeline of drug development, and have made a significant contribution to drug design and drug discovery \cite{macalino2015role}. AI-based drug design, one of the most important and rapidly evolving areas in CADD, has demonstrated great potential to revolutionized drug design and drug discovery \cite{mak2019artificial}. Impressive progresses have been made in various steps in virtural screening, including molecular docking \cite{ballester2010machine,khamis2015machine}, binding affinity prediction \cite{ain2015machine,jimenez2018k}, toxicity prediction \cite{mayr2016deeptox}, among others. Massive ambitious projects, between government agencies, researchers and biopharmaceutical companies, have been initialized \cite{smalley2017ai}. For instance, Pfizer is using IBM Watson to aid immuno-oncology drug discovery; AI models are used for personalization of treatment and early diagnosis in Obama's Cancer Moonshot initiative; Joint research consortium from Japanese companies and institutions use Japan's K supercomputer to ramp up drug discovery efficiency. AI-based drug design will bring about evolutional change with  more progress from chemical data, computational power, and learning algorithms  \cite{ekins2016next,chen2018rise}. Currently, one of the great challenges in AI-based drug design is molecular representation or featurization \cite{puzyn2010recent,lo2018machine}. In fact, the design of efficient molecular descriptors, is an open problem for the analysis of molecular structure-function relationship in materials sciences, chemistry and biology \cite{schutt2014represent,ramprasad2017machine,isayev2015materials,huan2015accelerated}. More than 5000 molecular descriptors have been proposed, which cover a variety of properties from molecular structural, physical, chemical and biological information \cite{puzyn2010recent,lo2018machine}. Recently, mathematical invariants from differential geometry and algebraic topology have been considered as molecular descriptors. Due to their higher level of abstraction and transferability, these advanced invariants based learning models have achieved great success in drug design \cite{cang:2017topologynet,cang:2018representability}.

As one of the fundamental concepts in different geometry, Ricci curvature characterizes the intrinsical properties of manifold surfaces \cite{jost2008riemannian,najman2017modern}. More specifically, Ricci curvature measures growth of volumes of distance balls, transportation distances between balls, divergence of geodesics, and meeting probabilities of coupled random walks \cite{samal2018comparative}. For two dimensional manifold, Ricci curvature reduces to the classical Gauss curvature. Ricci curvature based Ricci flow model is key to the proof of poincar\'{e} conjecture \cite{perelman2003ricci}. Discrete Ricci curvature forms, including Ollivier Ricci curvature (ORC)  \cite{ollivier2009ricci} and Forman Ricci curvature (FRC) \cite{forman2003bochner}, have been developed. In particularly, Chuang and Yau have done the pioneering works on graph or network based ORC  \cite{chung1996logarithmic,lin2011ricci}. The properties of ORCs on various graphs have been extensively studied \cite{bourne2018ollivier,samal2018comparative}. It has been found that ORC is ``related to" various graph invariants, ranging from local measures, such as node degree and clustering coefficient, to global measures, such as betweenness centrality and network connectivity \cite{ni2015ricci}. ORC has been used in various applications, such as internet topology  \cite{ni2015ricci}, community detection  \cite{ni2019community}, market fragility and systemic risk  \cite{sandhu2016ricci}, cancer networks  \cite{sandhu2015graph}, and brain structural connectivity  \cite{farooq2019network}.

Here we propose persistent Ricci curvature (PRC), in particular Ollivier persistent Ricci curvature (OPRC), based molecular representation, and apply them in AI-based drug design, for the first time. Essentially, molecular structures and interactions are modeled as graphs, on which ORCs can be calculated and used as geometric descriptors. To incorporate mutliscale structural and interactional information, a series of nested graphs at different scales are generated through a filtration process, the persistency and variation of ORCs during the filtration is defined as Ollivier persistent Ricci curvature. In particular, statistical or combinatorial properties of OPRC are considered as molecular descriptors or fingerpints, and further used in AI-based drug design models. We combine OPRC-based features with gradient boosting tree model (GBT), and test our OPRC-GBT model on three well-established protein-ligand binding affinity datasets, i.e., PDB-v2007, PDB-v2013, and PDB-v2016, from PBDbind databank. Our model can achieve the state-of-the-art results for all these datasets.

\section*{Results}

\subsection*{Ollivier Ricci curvature based molecular representation}

\begin{figure*}[ht!]
	\centering
	\includegraphics[width=0.7\linewidth]{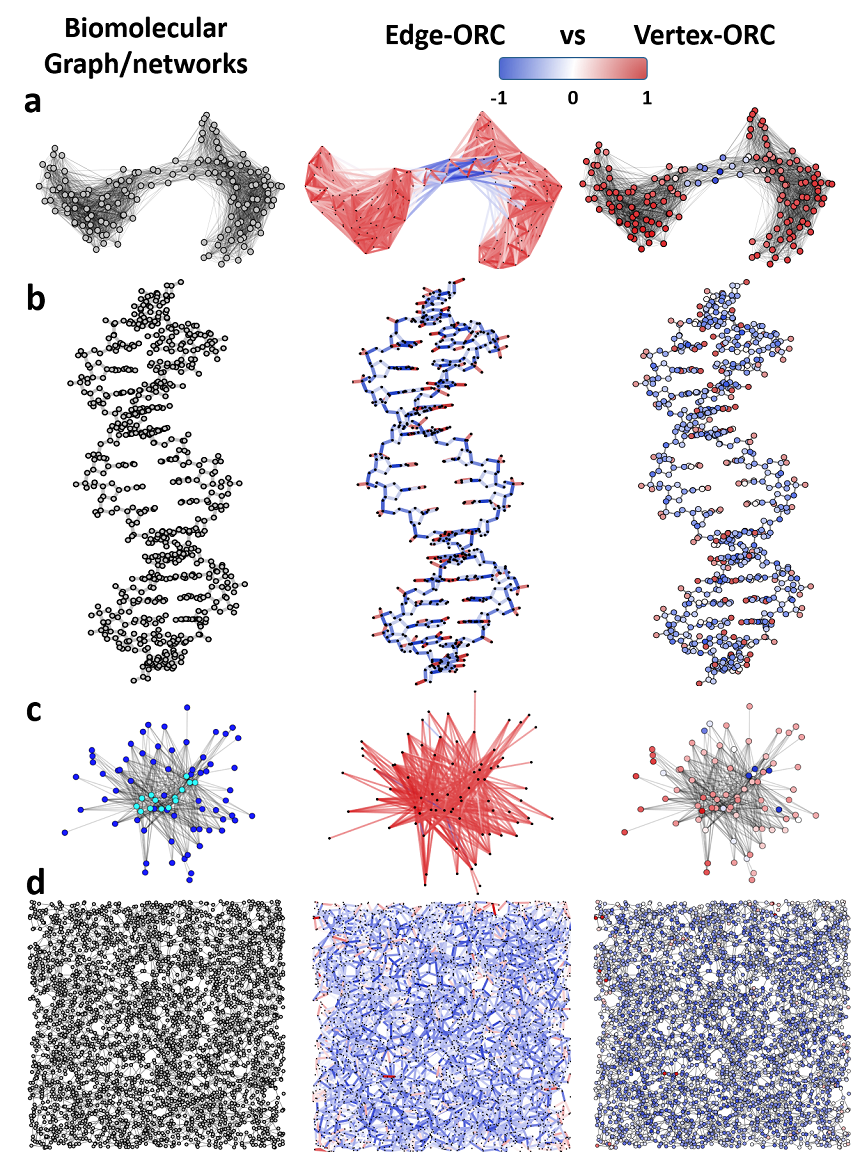}
	\caption{Illustration of edge and vertex ORCs for Calmodulin protein (4CLN), DNA (2M54), protein-ligand complex (1PXO), and hydrogen-bonding network of TMAO. 
{\bf a} Only $\text{C}_{\alpha}$ atoms are chosen in the graph representation and cutoff distance is 20\AA.
{\bf b} The molecular graph for DNA is generated with all non-hydrogen atoms and cutoff distance 2\AA. {\bf c} The molecular graph for protein (in blue) ligand (in cyan) complex is generated with oxygen atoms from protein and carbon atoms from ligand, and cutoff distance is 10\AA. {\bf d} The hydrogen bonding network for TMAO is generated with oxygen atoms from water molecules and cutoff distance 4\AA.}
	\label{fig:bio_orc}
\end{figure*}

As a discrete form of Ricci curvature on graph, Ollivier Ricci curvature measures the relative ``distance" between two respective neighborhoods of two vertices that form the edge. Roughly speaking, positive edge ORC means that there are strong connections (or short ``distance") between the two respective neighborhoods, and negative edge ORC indicates weak connections (or long ``distance"). A formal definition of ORC \cite{lin2011ricci,lin2010ricci,najman2017modern} can be found in Method. Figure \ref{fig:three_types} illustrates three basic types of quadratic surfaces, i.e., parabolic, hyperbolic and elliptic, and their related graph representations \cite{ni2019community}. For a complete graph, all edge ORCs are positive; for a tree graph, all edge ORCs are negative, except the ones on edges connected to leaf nodes; for a lattice graph, all edge ORCs are zero (see Methods). For all three situations, edge ORCs are consistent with Ricci curvatures in continuous definition. Mathematically, vertex ORC is defined as the average of edge ORCs from its adjacent edges \cite{ni2015ricci}. For isolated individual vertices, we define its vertex ORC as zero in this paper (see Method).

Here we employ ORC for the characterization of molecular structures and interactions at atomic level for the first time. Four typical molecular graphs/networks are generated from protein, DNA, protein-ligand interactions and hydrogen-bonding networks, by using $\text{C}_{\alpha}$ atoms, non-hydrogen atoms, both carbon (C) (from protein) and oxygen (O) (from ligand) atoms, and O atoms as vertices, respectively. Edges are formed in these molecular graphs, if Euclidean distances between the vertices are smaller than a certain cutoff distance. Figure \ref{fig:bio_orc} shows the four molecular representations, and their vertex and edge ORCs, which are represented by blue and red colors. Figure \ref{fig:bio_orc}{\bf a} is for a protein 4CLN, which has a hinge region in the middle of its structure and two cluster domains at its two ends. Both vertex and edge ORCs have negative values (blue color) in the hinge region and positive values in the cluster domains. DNA structure 2M54 is illustrated in Figure \ref{fig:bio_orc}{\bf b}. Other than the branching edges connected to the ``isolated" atom nodes, which have positive ORCs, all the other edges on DNA double-helix chains have negative ORCs. In Figure \ref{fig:bio_orc}{\bf c}, a bipartite graph is generated from protein-ligand complex 1PXO. Intermolecular interactions are represented as edges between protein atoms (in blue) and ligand atoms (in cyan). Since the graph is well-connected, almost all the vertex and edge ORCs are positive. Figure \ref{fig:bio_orc}{\bf d} demonstrates the hydrogen-bonding network generated with oxygen atoms as vertices and a cutoff distance 4\AA ~ \cite{xia2019persistent}. Under this short cutoff distance, an extremely sparse graph is formed and most of the vertex and edge ORCs are negative. From the above examples, it can be observed that edges with positive curvatures are commonly found in highly aggregated regions, such as clusters or communities, while edges with negative curvatures usually represent sparse connection regions, such as link, chain, or bridge regions.

As an intrinsical ``geometric" measurement, ORC can quantitatively characterize molecule structural and interactional properties. We consider two kinds of hydrogen-bonding networks from two osmolyte systems, i.e., TMAO and urea, at eight different concentrations from 1M to 8M  \cite{xia2019persistent}(See SI). Figures \ref{fig:osmolyte} {\bf a} and {\bf b} demonstrate the two equilibrium configurations from TMAO and urea, their edge and vertex ORCs and these ORC-based density distribution functions. Even though the two structures have the same number of water molecules (3000), their edge and vertex ORCs have very different distributions. It can be observed more clearly from their distribution functions (DFs), which have highly different peaks and shapes (see SI). Further, we consider ORC-based average DFs  for all eight concentrations, and illustrate the results in Figure \ref{fig:osmolyte} {\bf c}. We can see that two osmolyte systems show very different patterns in their ORC distributions. Dramatic variation between different concentrations can be seen clearly in TMAO ORC distribution functions. In contrast, only a slight variation is observed for urea ORC distribution functions. The results are consistent with our previous observation from persistent homology analysis \cite{xia2019persistent}.

\subsection*{Persistent Ricci curvature}

Biomolecules usually have hierarchial and multiscale structures, from various different levels ranging from atomic, residual, secondary structural, chain scale, to molecular monomer, and biomolecular complex. Both inter- and intra- biomolecular interactions are also of various scales, ranging from strong covalent bonds to related week interactions, such as hydrogen bonds, Van der Waals forces, electrostatic interactions, etc. Therefore, a multiscale representation is key to the characterization of biomolecular structures and interactions. Here, we use the filtration process, which is key component of persistent models, including persistent homology/cohomology  \cite{verri1993use,Edelsbrunner:2002,Zomorodian:2005}, persistent spectral  \cite{wang2020persistent,meng2020persistent} and persistent function \cite{bergomi2019beyond}, to generate a series of nested biomolecular graphs at different scales. A simple way to generate a filtration process is to set edge weight as a filtration parameter. As the increase (or decrease) of filtration value, edges with weights smaller (or larger) than the filtration value will consistently ``appear" to form a new graph. Since each filtration value characterizes structural or interactional information at a particular scale, a multiscale graph representation is obtained. Based on these nested graphs, Ricci curvatures can be systematically calculated. Among them, some Ricci curvature values last longer, while others change quickly during the filtration. We call this persistence and variation of Ricci curvatures during the filtration process as persistent Ricci curvature. In particularly, when we consider Ollivier Ricci curvature, we have Ollivier persistent Ricci curvature (OPRC).

\begin{figure*}[ht!]
	\centering
	\includegraphics[width=0.9\linewidth]{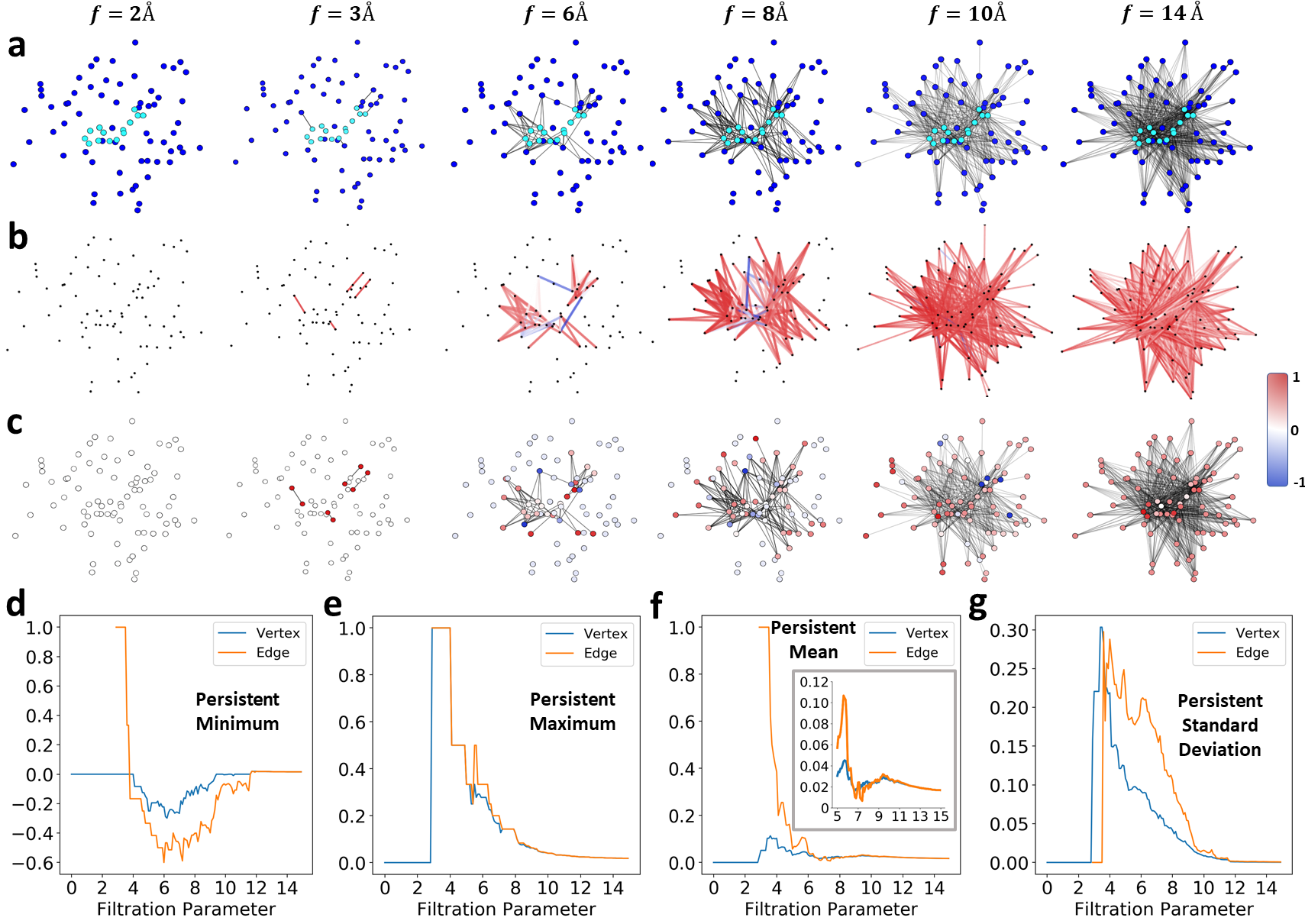}
	\caption{Illustration of edge ORCs ({\bf b}), vertex ORCs ({\bf c}), and four persistent attributes ({\bf d}-{\bf g}) for a bipartite graph based filtration process ({\bf a}). The bipartite graph is generated from a protein-ligand complex (1PXO) using O atoms from protein (in blue) and C atoms from ligand (in cyan). With the increase of filtration value, the bipartite graph inner connections (edges) increase consistently. At the same time, more and more positive ORCs (in red color) appear. At the very beginning of filtration (0 \AA~to 2.9 \AA), there are no edge ORCs and all vertex ORCs are zero. Therefore all four persistent attributes have only vertex values as 0.0. From filtration region 2.9 \AA ~ to 3.5 \AA, individual edges begin to appear. Their associated edge and vertex ORCs are all 1.0. Thus edge-ORC-based persistent minimum, maximum and mean all achieve their peak values 1.0 in this region. From region 3.5 \AA ~ to 10.0 \AA, some general trends for the four persistent attributes can be observed, even with some fluctuations. For both persistent minimum and mean, their vertex and edge values decrease until 7.0 \AA, and then increase to around 10.0\AA. In contrast, persistent maximum and standard deviation keep decreasing in this region. When the filtration value passes 10.0 \AA, all four persistent attributes begin to stabilized to certain constant values. }
	\label{fig:filtration_PA}
\end{figure*}

Figures \ref{fig:filtration_PA} {\bf a}-{\bf c} illustrate a filtration process from our element-specific interactive matrix (see Method) and the associated vertex and edge ORCs. It can be seen that at the early stages of filtration, negative ORCs (for both edges and vertices) are observed at link or bridge regions. As the filtration goes further, positive ORCs become more and more dominant. Finally, when the graph is well-connected, only positive ORCs can be observed. Note that different from all previous graph or network models, in which only a certain cutoff distance (i.e., single-scale) is considered,  we use a multiscale representation and focus on the variation and persistence of ORCs during the filtration process in our OPRC models.

\subsection*{OPRC-based machine learning model}

OPRC provides a new multiscale molecular representation. Based on them, a series of ORC-based persistent attributes or functions can be derived and used as molecular descriptors or fingerprints. For instance, we can consider ORC-based statistical properties, such as maximum, minimum, average and standard deviation, during the filtration process, and call them as ORC-based persistent maximum, persistent minimum, persistent mean, and persistent standard deviation, respectively. Figures \ref{fig:filtration_PA} {\bf d}-{\bf g} illustrate these four persistent attributes for the filtration process. Note that the filtration value is the cutoff distance. When filtration value is smaller than around 2.9 \AA, all vertices are ``isolated" and no edges are generated. In this case, there are no edge ORCs and all vertex ORCs equal to zero. For the four persistent attributes, there are no well-defined edge values and their vertex values all equal to zero. From the region around 2.9 \AA ~ to 3.5 \AA, individual edges, i.e., no two edges share same vertex, begin to appear. For these individual edges and their associated vertices, their ORCs (for both edge and vertex) all equal to 1.0. Therefore, edge-ORC-based persistent minimum, persistent maximum and persistent mean all achieve the largest value 1.0, and edge-ORC-based persistent standard deviation is zero. At the same time, vertex-ORC-based persistent minimum value is still 0.0, persistent maximum jumps to 1.0, and both persistent mean and standard deviation consistently increase. From the region around 3.5 \AA~to 10.0 \AA, more edges are generated and graph structures become more complicated. Both edge and vertex ORCs have positive and negative values. For persistent minimum, both its vertex and edge values are negative. They keep decreasing to around 7.0 \AA~ and then begin to increase until reach 0.0 at around 10 \AA~ (for vertex) and 12 \AA~(for edge). Persistent mean has a similar pattern to persistent minimum, both edge and vertex values decrease until around 7.0 \AA~ and then increase to reach the peak at around 10 \AA. For all the other two persistent attributes, both their vertex and edge values are in the same general trend of decreasing, even though there are certain fluctuations. When the filtration value is larger than 10 \AA, the molecular graph becomes more and more ``well-connected". Almost all the vertex and edge ORCs are positive, or at least non-negative. With the increase of filtration value, a fully-connected bipartite graph (with the protein partite set of 72 O atoms and the ligand partite set of 14 C atoms) will be generated. Both vertex and edge ORCs will converge to $\frac{1}{72} \approx$ 0.0139 (see Eq.(\ref{eq:bipartite_curvature}) in Method ). Similarly, persistent minimum, persistent maximum, and persistent mean, for both vertex and edge ORCs, all converge to 0.0139, while both persistent standard deviations converge to 0.00.


Other than these simple statistical properties, we can also consider combinatorial formulas  \cite{puzyn2010recent}, such as moments, Wiener-index, etc, to generate more complicated persistent attributes functions (see Methods). Molecular descriptors can be derived from the discretization of these OPRC-based persistent attributes. More specifically, we can discretize the entire filtration region into equal-sized bins. Persistent attributes from each bin can be combined together to form a long feature vector. Note that these feature vectors are of the same unit and highly abstract. They can be combined with machine learning models, in particular, random forest and gradient boosting tree.

\subsection*{OPRC-based machine learning model for drug design}

\begin{figure*}[ht!]
	\centering
	\includegraphics[width=.8\linewidth]{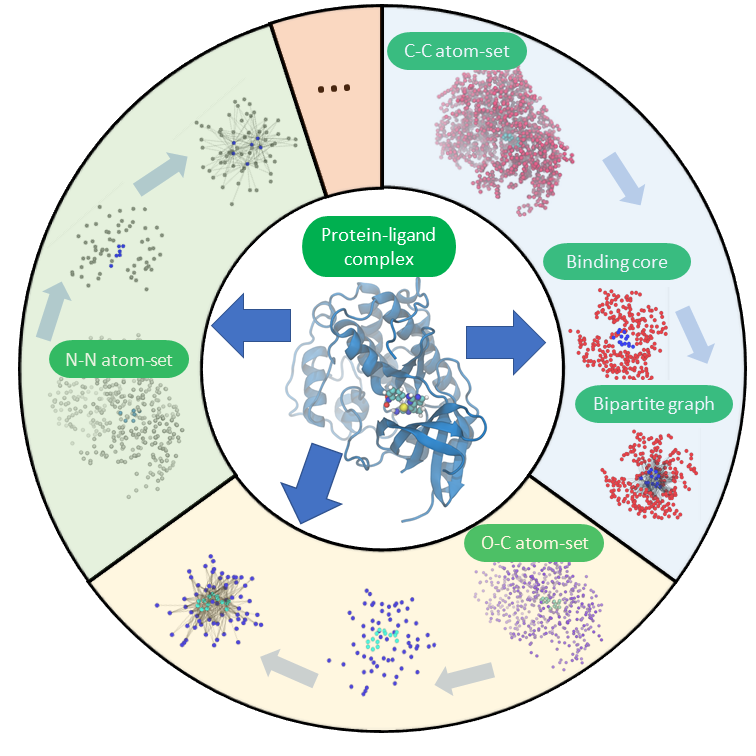}
	\caption{Generation of the 36 types of element-specific bipartite graphs. The protein-ligand complex (PDBID:1PXO) is decomposed into its atom-atom combinations (outer circle): N-N(left), O-C(bottom) and C-C(right).}
	\label{fig:ML_flow_own}
\end{figure*}

Here we examine the performance of our OPRC-based learning models for drug design. In particular, we study the protein-ligand binding affinity, which is of key importance to drug design \cite{li2015improving}. We consider three most commonly-used protein-ligand databases, including PDBbind-v2007, PDBbind-v2013 and PDBbind-v2016  \cite{PDBBind:2015}. The details of the datasets can be found in Table \ref{tab:PDBbind}.

In our protein-ligand interaction model, 36 types of bipartite graphs are generated based on element-specific interactive distance matrixes (ES-IDM) \cite{cang:2017topologynet,cang:2018representability}, as illustrated in Figure \ref{fig:ML_flow_own}. For each type of bipartite graphs, we consider 10 different persistent attributes, and generate the corresponding molecular descriptors from them (see Methods). The feature size of our OPRC based molecular descriptor is 108000 = 36(atom set combinations)*150(filtration values)*10 (persistent attributes)* 2(edge ORC and vertex ORC). To take into consideration of ligand properties, we incorporate molecular descriptors from ligands. Similarly, we decompose a ligand structure into 5 sets of atom combinations, construct related ligand graphs and extract molecular features from their persistent attributes (see Methods). The total feature size for ligand is 15000= 5(atom set combinations)*150(filtration values)*10 (persistent attributes)* 2(edge ORC and vertex ORC).

\begin{figure*}[ht!]
	\centering
	\includegraphics[width=0.8\linewidth]{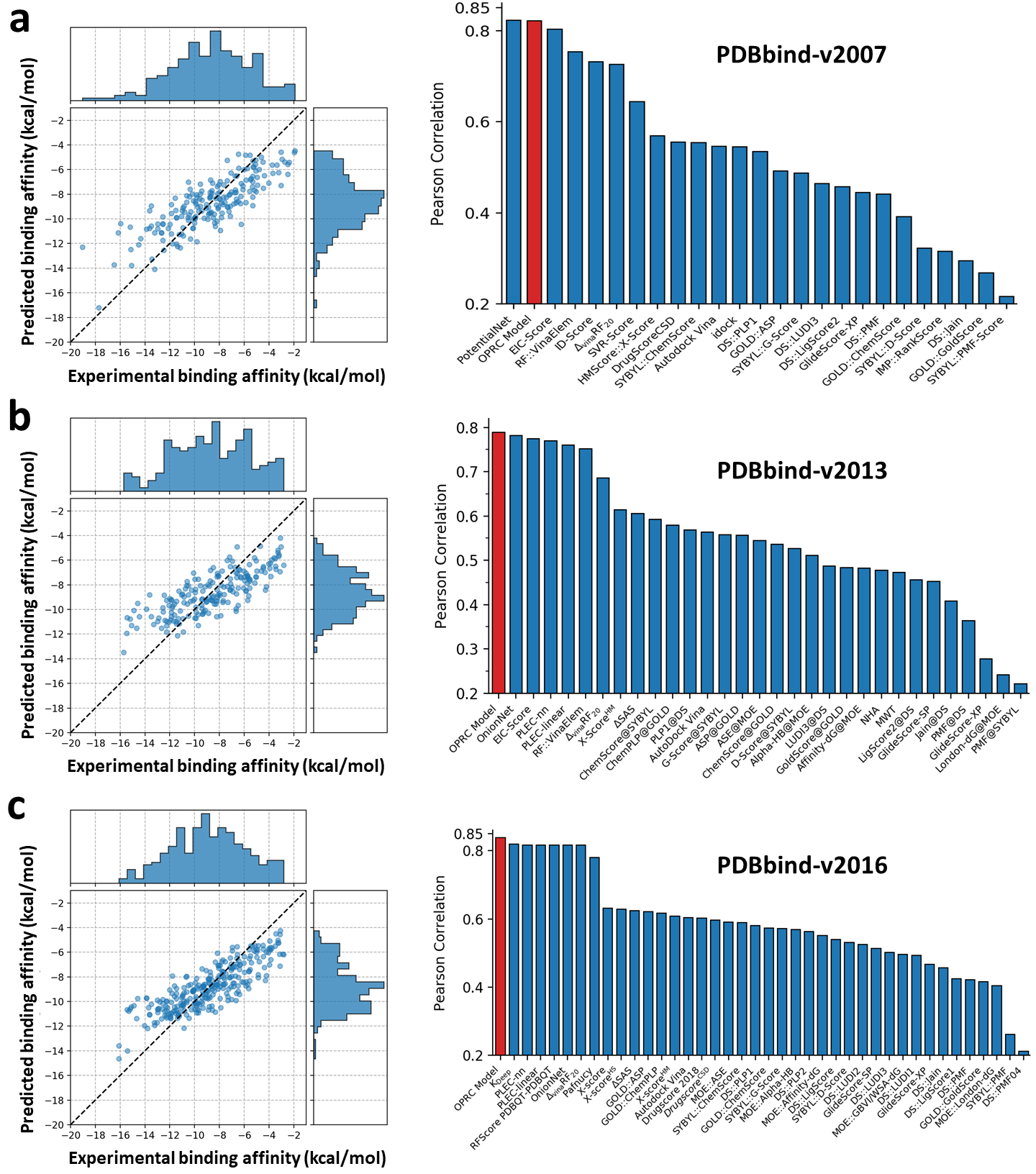}
	\caption{The comparison of our OPRC-GBT model with all traditional-molecular-descriptor based machine learning models \cite{liu2015classification,li2015improving,wojcikowski2019development,jimenez2018k,stepniewska2018development,su2018comparative,afifi2018improving}, as far as we know, for protein-ligand binding affinity prediction. The left subgraphs are results from our OPRC-GBT models with combined molecular features (\textbf{Pro-Lig + Lig}). The right subfigures are comparison of PCCs values on the three datasets, i.e., PDBbind-v2007 ({\bf a}), PDBbind-v2013 ({\bf b}) and PDBbind-v2016 ({\bf c}). }
	\label{fig:corr_plot}
\end{figure*}

Due to the large-sized molecular descriptor, we use decision-tree-based machine learning models, in particular, gradient boosting tree (GBT). To measure the predicted binding affinities, the Pearson correlation coefficient (PCC) and root mean square error (RMSE) are used as evaluation metrics. Our GBT model has been repeated for 10 times and median PCCs and RMSEs for test set are used as our final results. The detailed setting of GBT can be found in Table \ref{tab:GBT-hyperparam}. Table \ref{tab:PerORC} illustrates all the prediction results using features from protein-ligand complex (\textbf{Pro-Lig}), and the combined features from protein-ligand complexes with ligands (\textbf{Pro-Lig + Lig}), for the test (core) datasets of PDBbind databases. For OPRC-GBT with molecular features from protein-ligand interaction (\textbf{Pro-Lig}),  the average PCC is 0.812 and RMSE is 1.906 kcal/mol. For OPRC-GBT with combined molecular features  (\textbf{Pro-Lig+Lig}), the average PCC has a slight increase to 0.816, and RMSE has a slight decrease to 1.891 kcal/mol. To have better evaluation of our results,  we have extensively compared with the results from all existing machine learning models using traditional molecular descriptors and fingerprints \cite{liu2015classification,li2015improving,wojcikowski2019development,jimenez2018k,stepniewska2018development,su2018comparative,afifi2018improving}, as far as we know. Figure \ref{fig:corr_plot} shows the comparison results for PCCs. It can be seen that our model can achieve the state-of-the-art results.

\begin{table}[H]
	\centering
\caption{The PCCs and RMSE (kcal/mol) for our OPRC-GBT models with molecular features from protein-ligand complex(\textbf{Pro-Lig}) and combined protein-ligand complex and ligand (\textbf{Pro-Lig + Lig}). The results are based on the test (core) datasets of PDBbind-v2007, PDBbind-v2013 and PDBbind-v2016.}
	\begin{tabular}{|c|c|c|}
		\hline
		\textbf{}     & \textbf{Pro-Lig}  & \textbf{Pro-Lig + Lig} \\ \hline
		PDBbind-v2007 & 0.820 (1.935) & 0.821  (1.926)   \\ \hline
		PDBbind-v2013 & 0.781 (2.035) & 0.789  (2.010)  \\ \hline
		PDBbind-v2016 & 0.835 (1.748) & 0.838  (1.736)   \\ \hline
            Average   & 0.812 (1.906) &  0.816 (1.891)   \\ \hline
	\end{tabular}
	\label{tab:PerORC}
\end{table}

\section*{Discussion}
Efficient molecular descriptors or fingerprints are essential not only to artificial intelligence (AI) based drug design, and also all the learning models in materials, chemical and biological data analysis. Unlike image, text, video, and audio data, molecular data from material, chemistry and biology, have much complicated three-dimensional structures, as well as physical and chemical properties. A systematical generation of descriptors that characterizes the molecular intrinsic properties can directly determinate the performance of learning models. As a central concept in geometry, curvature is of fundamental importance to differential geometry and Riemannian geometry. Among the various curvature definitions, Ricci curvature can characterize the intrinsical surface properties, and their discrete forms (Ollivier Ricci curvature and Forman Ricci curvature) provide a unique ``geometric" measurement of graphs and networks. Here we use Ollivier Ricci curvature to characterize molecular structures and interactions, and propose persistent Ricci curvature, in particular, Ollivier persistent Ricci curvature, for machine learning based drug design. Essentially, persistent attributes based molecular descriptors from our OPRC, can preserve the multiscale structural and interactional information, thus can significantly boost the performance of learning models.

The recent great success of persistent homology, persistent spectral models and our persistent Ricci curvature models, has shown that advanced mathematical tools from algebraic topology, computational topology, differential geometry, conformational geometry, etc, can play an  important role in structural representation, thus provide an efficient featurization or feature engineering. Their combination with machine learning models can significantly boost the performance of learning models. It is worthy mentioning that graph neural network and geometric deep learning have also shown some great promising in automatic feature generation from molecular structures.

\section*{Method}

\subsection*{Ollivier Ricci curvature}
Mathematically, Ricci curvature measures manifold local intrinsic properties, such as growth of volumes of distance balls, transportation distances between balls, divergence of geodesics, and meeting probabilities of coupled random walks \cite{samal2018comparative}. Ollivier Ricci curvature is a discrete formula of Ricci curvature proposed on metric spaces  \cite{ollivier2007ricci,ollivier2009ricci}, and further applied to graphs \cite{lin2011ricci,lin2010ricci}. ORC is defined on an edge. It measures the difference between the edge ``distance" (or length of edge) and transportation distance of the two neighborhoods from the-edge-related two vertices. More specifically, we can denote a graph as $G=(V, E)$ with vertex set $V=\{x_i;i=1,2,..,n\}$ and edge set $E=\{e(x_{i},x_{j}); 1 \leq i<j \leq n\}$ ($n$ is total number of vertices). If two vertices $x$ and $y$ share a common edge $e(x,y)$, we call these two vertices as neighbors and denote them as $x\sim y$. All the neighboring vertices of $x$ form a set $\Gamma_x$, and their total number is denoted $k_x$, i.e., the degree of $x$. A vertex $x$ related probability measure $m_x$ is defined as
\begin{equation}\label{eq:probability}
	m_x (x_i)= \begin{cases}
	\alpha & \mbox{if}\ x_i = x \\
	(1-\alpha)/k_x & \mbox{if}\ x_i \in \Gamma_x \\
	0 & \mbox{otherwise.}
	\end{cases}
\end{equation}
Here vertex $x_i \in V$ and parameter $\alpha\in [0,1]$. The probability measure $m_x$ is a discrete function that defines on the graph vertices, and has nonzero value only on vertex $x$ and its neighbors. Note that different  $\alpha$ values will result in different ORCs. Usually $\alpha=0.5$ is considered as a good approximation. In this paper, we use $\alpha=0.5$ in all our calculations.

For two probability measures $m_x$ and $m_y$, a transportation plan from $m_x$ to $m_y$ is a measure $\xi\in\prod(m_x,m_y)$ that is mass-preserving, i.e. $\sum_{y_j \in V} \xi(x,y_j) = m_x$ and $\sum_{x_i \in V} \xi(x_i,y) = m_y$, where $\xi(x_i,y_j)$ represents the amount of mass travelling from $x_i$ to $y_j$. The $L_1$ Wasserstein distance between $m_x$ and $m_y$, denoted by $W_1(m_x,m_y)$, is the minimum average travelling distance that can be achieved by any transportation plan,
	\begin{equation} \nonumber
	W_1(m_x, m_y) = \inf_{\xi} \sum_{x_i \in V}\sum_{y_j \in V} \;d(x_i,y_j)\xi(x_i,y_j).
	\end{equation}
If $x \sim y$, the Olliver-Ricci curvature along the edge between vertex $x$ and $y$ is defined as
	\begin{equation} \nonumber
	c(x,y) := 1- \frac{W_1(m_x, m_y)}{d(x,y)},
	\end{equation}
where $d(x,y)$ is the distance between $x$ and $y$ (edge distance or edge length) on the graph $G$. Note that the above definition of ORC is only for edges, that is edge ORC. Vertex ORC is defined as the average of edge ORCs from its adjacent edges. Moreover, we define vertex ORCs as zero in this paper, for all isolated individual vertices, i.e., no two vertices share a same edge.

The edge ORC is computed via the optimizing Wasserstein distance using linear programming. For two vertices $x$ and $y$ and associated probability measures $m_x$ and $m_y$, let $\rho (x_i,y_j)$ be the proportion of ``mass" ($m_x$ value) transported from vertex $x_i$ to vertex $y_j$. The objective function is to minimize the total transportation distance between $m_x$ and $m_y$ (on the graph $G$), while the total ``mass" is preserved. The linear programming can be written as follows,
\begin{align} \nonumber
\min &\sum_{y_j \in V}\sum_{x_i \in V} d(x_i, y_j) \rho (x_i, y_j) m_{x} (x_i),\\ \nonumber
\text{such that } &\sum_{y_j \in V} \rho (x_i, y_j) = 1,  \quad 0\le \rho (x_i, y_j) \le 1,\\ \nonumber
& \sum_{x_i \in V} \rho(x_i, y_j) m_x(x_i) = m_y(y_j).
\end{align}

Figure \ref{fig:three_types} illustrates the edge ORC on three basic types of graphs, i.e., tree, grid, and complete graph, which correspond to three types of quadratic surfaces, i.e., parabolic, hyperbolic, elliptic \cite{ni2019community}. For a complete graph in Figure \ref{fig:three_types} {\bf a}, the edge ORC on $e(x,y)$ can be explicitly expressed as,
	\begin{equation} \nonumber
	c(x,y) = \frac{n}{2(n-1)},
	\end{equation}
where $n$ is the total number of vertices. Note that $\alpha=0.5$ is always used in our calculation.

For an infinitely-sized grid graph in Figure \ref{fig:three_types} {\bf b},  all edge ORCs equal to zero. This is due to the fact that the cost of moving $m_x$ to $m_y$ is equal to $d(x,y)$, which means $W_1(m_x, m_y)=d(x,y)$.
For a tree graph in Figure \ref{fig:three_types} {\bf c}, the edge ORC for $e(x,y)$ can be explicitly expressed as,
	\begin{equation} \nonumber
	c(x,y)=\frac{1}{k_x}+\frac{1}{k_y} - 1,
	\end{equation}
where $k_x$ and $k_y$ are node degrees for vertices $x$ and $y$.

For a complete bipartite graph $K_{m,n}$ with $m$ vertices at one { partite set} and $n$ vertices at the other { partite set}, the edge ORC for $e(x,y)$ is,
\begin{equation} \label{eq:bipartite_curvature}
c(x,y) = \min\bigg\{\frac{1}{m}, \frac{1}{n}\bigg\}.
\end{equation}
A detailed derivation of the above results can be found in SI.

\subsection*{Ollivier persistent Ricci curvature}

A well-defined filtration process is essential to our persistent OPRC model. The idea of filtration is originated from persistent homology and can be traced back to Morse theory. Mathematically, with the monotonic variation (increase or decrease) of filtration value, a series of nested graphs $\{ G_i \}$ can be generated as follows,
\begin{align} \nonumber
G_0 \subseteq G_1 \subseteq G_2  ... \subseteq G_{N},
\end{align}
here $i$-th graph $G_i$ is generated at a certain filtration value $f_i$, and $G_N$ is usually a complete graph at the end of filtration. The number of edges are consistently increasing during filtration process. In contrast, vertices are usually kept unchanged in real application, even though theoretically their total number can still increase. A proper filtration parameter directly determines the performance of OPRC. For weighted networks or graphs, filtration value can be chosen as edge weight. At each filtration value $f_i$, a subgraph $G_i$ composed of all edges with weight less or equal to $f_i$ is formed. Figure \ref{fig:filtration_PA} illustrates a bipartite graph based filtration process.

In our OPRC model, edge and vertex ORCs can be calculated for the graph series $\{ G_i \}$. The persistence and variation of ORCs are defined as OPRC for the graph series $\{ G_i \}$. Note that it is nontrivial to visualize OPRC or apply it directly to machine learning models. In this way, we propose a set of persistent attributes from statistical and combinatorial properties \cite{puzyn2010recent} of ORCs. For instance, if we denote a set of (edge or vertex) ORC as $\{c_j; j=1,2,...,N \}$, we can consider maximum, minimum, summary, average, standard deviation, $k$-th moment $\sum_j^N c_j^k$, $k$-th absolute deviation $\sum_j^N |c_j - \sum_{j=1} c_j|^k$, etc, for both edge ORC and vertex ORCs in a filtration process.

In our molecular description, we consider 10 OPRC descriptors as follows,
\begin{itemize}
\item  $\min \{c_1, c_2, \cdots, c_N\}$
\item  $\max\{c_1, c_2, \cdots, c_N\}$
\item  $\frac{1}{N} \sum_{i=1}^{N} c_j$
\item  $\sqrt{\frac{1}{N-1}\sum_{j=1}^N (c_j - \bar{c})^2 }$
\item  $\sum_{j=1}^{N} c_j\mathbbm{1}_{\{c_j > 0\}}$  ($\mathbbm{1}_{c_j > 0}$ is indicator function). 
\item  $\sum_{j=1}^N |c_j - \bar{c}|$
\item  $\sum_{j=1}^{N} c_j^2$
\item $\sum_{j=1}^{N} c_j^2\mathbbm{1}_{\{c_j > 0\}}$
\item  $\sum_{j=1}^N |c_j - \bar{c}|^3$
\item  $\log\bigg(N\sum_{j=1, c_j>0}^{N^+} \frac{1}{c_j} + 1\bigg)$ ($N^+$ is the total positive element number).
\end{itemize}
Here curvature quasi-Wiener index $\log\bigg(N\sum_{j=1}^{N^+} \frac{1}{c_j} + 1\bigg)$ is a variant of the quasi-Wiener index  \cite{puzyn2010recent}. The logarithmic function is empolyed to scale down the values, as ORCs are always in the region $[-1,1]$.

\subsection*{Protein-ligand binding affinity prediction with OPRC}
The protein-ligand binding affinity is used to characterize the interactions between protein and ligand. In our OPRC models, we use the bipartite graph representation generated from the element-specific interactive distance matrix (ES-IDM), to describe the protein-ligand interactions at atomic level \cite{cang:2017topologynet,cang:2018representability}. More specifically, protein structures are decomposed into 4 atom sets made of { C, N, O and S}, respectively. Ligands are decomposed into 9 atom sets of { C, N, O, S, P, F, Cl, Br and I}, respectively. In this way, there are totally 36 element-specific interaction types for protein-ligand complexes, and they are characterized by 36 ES-IDMs. We denote atom coordinate as $\mathbf{r}$, and atom-set from protein and ligand as $\mathbf{R}_P=\{ \mathbf{r_1}, \mathbf{r_2},..., \mathbf{r_{N_P}}\}$ and $\mathbf{R}_L=\{ \mathbf{r_1}, \mathbf{r_2},..., \mathbf{r_{N_L}}\}$, respectively. An ES-IDM $M$ of size $(N_P+N_L)\times(N_P+N_L)$ is derived from the two atom sets $\mathbf{R}_P$ and $\mathbf{R}_L$ as follows,
	\[ M(m_i,m_j) = \begin{cases}  \nonumber
	 { \scriptstyle \|\mathbf{r}_i - \mathbf{r}_j\|}, & { \scriptstyle\mathbf{r}_i \in \mathbf{R}_P, \mathbf{r}_j \in \mathbf{R}_L ~\text{or}~ \mathbf{r}_i \in \mathbf{R}_L, \mathbf{r}_j \in \mathbf{R}_P} \\
	\infty, & \text{otherwise}.
	\end{cases} \]
where $\|\mathbf{r}_i - \mathbf{r}_j\|$ is the Euclidean distance, and $m_i$ and $m_j$ are the indexes of protein atom $\mathbf{r}_i$ and ligand atom $\mathbf{r}_j$ in matrix $M$, respectively.

Computationally, we only consider the binding core region or hot spot, which is composed of atoms within $10$ \AA \space distance  from the ligand. This is because protein structures are usually much larger than ligand structures. The binding core complex, i.e., protein binding core with ligand, is then decomposed into the 36 element-specific combination sets (as illustrated in Figure \ref{fig:ML_flow_own}), which will generate 36 ES-IDMs.  The filtration parameter is chosen as the cutoff distance. The filtration region from 0 to 15 \AA\space and a grid size 0.1 \AA\space are considered in our model. At each filtration value, edge and vertex ORCs are calculated. A total 10 persistent attributes as stated above are used. In this way, the size of our molecular descriptor is 108000 = 36(atom set combinations)*150(filtration values)*10 (persistent attributes)* 2(edge ORC and vertex ORC).

Molecular descriptors from ligand structures have been found to play a role in binding affinity prediction. We consider the following 5 atom set combinations in our ligand feature generation: $\big\{$\{C\}, \{C, N\}, \{C, O\}, \{C, N, O\}, \{C, N, O, F, P, Cl, Br, I\}$\big\}$. A distance matrix can be constructed for each atom combination as follows,
\[ M^L(i,j) = \begin{cases}  \nonumber
	\|\mathbf{r}_i - \mathbf{r}_j\|, & \mathbf{r}_i \in \mathbf{R}_L, \mathbf{r}_j \in \mathbf{R}_L, i \neq j\\
	0, & \mathbf{r}_i \in \mathbf{R}_L, \mathbf{r}_j \in \mathbf{R}_L, i =j.
	\end{cases} \]
where $\|\mathbf{r}_i - \mathbf{r}_j\|$ is the Euclidean distance. From the distance matrixes, ligand graphs and cutoff distance based filtration process can be generated. Ligand features can be obtained from our OPRC of ligand graphs. The same 10 persistent attributes are used in molecular feature generation. Hence, a ligand feature vector is of size  15000=5(atom set combinations)*150(filtration values)*10 (persistent attributes)* 2(edge ORC and vertex ORC). Note that the concatenation of \textbf{Pro-Lig} and \textbf{Lig} would have a feature size of 123000.

\section*{Code and Data Availability}
The computation of Ollivier Ricci curvature for a graph network is obtained using the GraphRicciCurvature library from
\href{https://github.com/saibalmars/GraphRicciCurvature}{https://github.com/saibalmars/GraphRicciCurvature}. The PDBbind databases were obtained from \href{http://pdbbind.org.cn}{http://pdbbind.org.cn}. The
codes implemented for the HBNs and PerORC models can be found in \href{https://github.com/ExpectozJJ/Persistent-Ollivier-Ricci-Curvature}{https://github.com/ExpectozJJ/Persistent-Ollivier-Ricci-Curvature}. Additional data or code would be available upon reasonable request.


\begin{thebibliography}{10}

\bibitem{macalino2015role}
Macalino SJY, Gosu V, Hong S, Choi S (2015) Role of computer-aided drug design
  in modern drug discovery.
\newblock {\em Archives of pharmacal research} 38(9):1686--1701.

\bibitem{mak2019artificial}
Mak KK, Pichika MR (2019) Artificial intelligence in drug development: present
  status and future prospects.
\newblock {\em Drug discovery today} 24(3):773--780.

\bibitem{ballester2010machine}
Ballester PJ, Mitchell JB (2010) A machine learning approach to predicting
  protein--ligand binding affinity with applications to molecular docking.
\newblock {\em Bioinformatics} 26(9):1169--1175.

\bibitem{khamis2015machine}
Khamis MA, Gomaa W, Ahmed WF (2015) Machine learning in computational docking.
\newblock {\em Artificial intelligence in medicine} 63(3):135--152.

\bibitem{ain2015machine}
Ain QU, Aleksandrova A, Roessler FD, Ballester PJ (2015) Machine-learning
  scoring functions to improve structure-based binding affinity prediction and
  virtual screening.
\newblock {\em Wiley Interdisciplinary Reviews: Computational Molecular
  Science} 5(6):405--424.

\bibitem{jimenez2018k}
Jim{\'e}nez J, Skalic M, Martinez-Rosell G, De~Fabritiis G (2018) K$_{DEEP}$:
  Protein--ligand absolute binding affinity prediction via {3D}-convolutional
  neural networks.
\newblock {\em Journal of chemical information and modeling} 58(2):287--296.

\bibitem{mayr2016deeptox}
Mayr A, Klambauer G, Unterthiner T, Hochreiter S (2016) Deeptox: toxicity
  prediction using deep learning.
\newblock {\em Frontiers in Environmental Science} 3:80.

\bibitem{smalley2017ai}
Smalley E (2017) {AI}-powered drug discovery captures pharma interest.
\newblock {\em Nature} (35):604--605.

\bibitem{ekins2016next}
Ekins S (2016) The next era: deep learning in pharmaceutical research.
\newblock {\em Pharmaceutical research} 33(11):2594--2603.

\bibitem{chen2018rise}
Chen H, Engkvist O, Wang Y, Olivecrona M, Blaschke T (2018) The rise of deep
  learning in drug discovery.
\newblock {\em Drug discovery today} 23(6):1241--1250.

\bibitem{puzyn2010recent}
Puzyn T, Leszczynski J, Cronin MT (2010) {\em Recent advances in {QSAR}
  studies: methods and applications}.
\newblock (Springer Science \& Business Media) Vol.{}~8.

\bibitem{lo2018machine}
Lo YC, Rensi SE, Torng W, Altman RB (2018) Machine learning in chemoinformatics
  and drug discovery.
\newblock {\em Drug discovery today} 23(8):1538--1546.

\bibitem{schutt2014represent}
Sch{\"u}tt KT, Glawe H, Brockherde F, Sanna A, M{\"u}ller, K. R.and~Gross EKU
  (2014) How to represent crystal structures for machine learning: Towards fast
  prediction of electronic properties.
\newblock {\em Physical Review B} 89(20):205118.

\bibitem{ramprasad2017machine}
Ramprasad R, Batra R, Pilania G, Mannodi-Kanakkithodi A, Kim C (2017) Machine
  learning in materials informatics: recent applications and prospects.
\newblock {\em npj Computational Materials} 3(1):54.

\bibitem{isayev2015materials}
Isayev O, et~al. (2015) Materials cartography: representing and mining
  materials space using structural and electronic fingerprints.
\newblock {\em Chemistry of Materials} 27(3):735--743.

\bibitem{huan2015accelerated}
Huan TD, Mannodi-Kanakkithodi A, Ramprasad R (2015) Accelerated materials
  property predictions and design using motif-based fingerprints.
\newblock {\em Physical Review B} 92(1):014106.

\bibitem{cang:2017topologynet}
Cang ZX, Wei GW (2017) {TopologyNet: Topology} based deep convolutional and
  multi-task neural networks for biomolecular property predictions.
\newblock {\em PLOS Computational Biology} 13(7):e1005690.

\bibitem{cang:2018representability}
Cang ZX, Mu L, Wei GW (2018) Representability of algebraic topology for
  biomolecules in machine learning based scoring and virtual screening.
\newblock {\em PLoS computational biology} 14(1):e1005929.

\bibitem{jost2008riemannian}
Jost J, Jost J (2008) {\em Riemannian geometry and geometric analysis}.
\newblock (Springer) Vol.{} 42005.

\bibitem{najman2017modern}
Najman L, Romon P (2017) {\em Modern approaches to discrete curvature}.
\newblock (Springer) Vol.{} 2184.

\bibitem{samal2018comparative}
Samal A, et~al. (2018) Comparative analysis of two discretizations of {Ricci}
  curvature for complex networks.
\newblock {\em Scientific reports} 8(1):1--16.

\bibitem{perelman2003ricci}
Perelman G (2003) Ricci flow with surgery on three-manifolds.
\newblock {\em arXiv preprint math/0303109}.

\bibitem{ollivier2009ricci}
Ollivier Y (2009) Ricci curvature of markov chains on metric spaces.
\newblock {\em Journal of Functional Analysis} 256(3):810--864.

\bibitem{forman2003bochner}
Forman R (2003) Bochner's method for cell complexes and combinatorial {Ricci}
  curvature.
\newblock {\em Discrete and Computational Geometry} 29(3):323--374.

\bibitem{chung1996logarithmic}
Chung FR, Yau ST (1996) Logarithmic harnack inequalities.
\newblock {\em Mathematical Research Letters} 3(6):793--812.

\bibitem{lin2011ricci}
Lin Y, Lu L, Yau ST (2011) Ricci curvature of graphs.
\newblock {\em Tohoku Mathematical Journal, Second Series} 63(4):605--627.

\bibitem{bourne2018ollivier}
Bourne DP, Cushing D, Liu S, Munch F, Peyerimhoff N (2018) Ollivier-{Ricci}
  idleness functions of graphs.
\newblock {\em SIAM Journal on Discrete Mathematics} 32(2):1408--1424.

\bibitem{ni2015ricci}
Ni CC, Lin YY, Gao J, Gu XD, Saucan E (2015) Ricci curvature of the internet
  topology in {\em 2015 IEEE Conference on Computer Communications (INFOCOM)}.
\newblock (IEEE), pp. 2758--2766.

\bibitem{ni2019community}
Ni CC, Lin YY, Luo F, Gao J (2019) Community detection on networks with {Ricci}
  flow.
\newblock {\em Scientific reports} 9(1):1--12.

\bibitem{sandhu2016ricci}
Sandhu RS, Georgiou TT, Tannenbaum AR (2016) Ricci curvature: An economic
  indicator for market fragility and systemic risk.
\newblock {\em Science advances} 2(5):e1501495.

\bibitem{sandhu2015graph}
Sandhu R, et~al. (2015) Graph curvature for differentiating cancer networks.
\newblock {\em Scientific reports} 5(1):1--13.

\bibitem{farooq2019network}
Farooq H, Chen Y, Georgiou TT, Tannenbaum A, Lenglet C (2019) Network curvature
  as a hallmark of brain structural connectivity.
\newblock {\em Nature communications} 10(1):1--11.

\bibitem{lin2010ricci}
Lin Y, Yau ST (2010) Ricci curvature and eigenvalue estimate on locally finite
  graphs.
\newblock {\em Mathematical research letters} 17(2):343--356.

\bibitem{xia2019persistent}
Xia KL, Anand DV, Saxena S, Mu YG (2019) Persistent homology analysis of
  osmolyte molecular aggregation and their hydrogen-bonding networks.
\newblock {\em Physical Chemistry Chemical Physics} 21(37):21038--21048.

\bibitem{verri1993use}
Verri A, Uras C, Frosini P, Ferri M (1993) On the use of size functions for
  shape analysis.
\newblock {\em Biological cybernetics} 70(2):99--107.

\bibitem{Edelsbrunner:2002}
Edelsbrunner H, Letscher D, Zomorodian A (2002) Topological persistence and
  simplification.
\newblock {\em Discrete Comput. Geom.} 28(4):511--533.

\bibitem{Zomorodian:2005}
Zomorodian A, Carlsson G (2005) Computing persistent homology.
\newblock {\em Discrete Comput. Geom.} 33(2):249--274.

\bibitem{wang2020persistent}
Wang R, Nguyen DD, Wei GW (2020) Persistent spectral graph.
\newblock {\em International Journal for Numerical Methods in Biomedical
  Engineering} p. e3376.

\bibitem{meng2020persistent}
Meng ZY, Xia KL (2020) Persistent spectral based machine learning ({PerSpect
  ML}) for drug design.
\newblock {\em arXiv preprint arXiv:2002.00582}.

\bibitem{bergomi2019beyond}
Bergomi MG, Ferri M, Vertechi P, Zuffi L (2019) Beyond topological persistence:
  Starting from networks.
\newblock {\em arXiv preprint arXiv:1901.08051}.

\bibitem{liu2015classification}
Liu J, Wang RX (2015) Classification of current scoring functions.
\newblock {\em Journal of chemical information and modeling} 55(3):475--482.

\bibitem{li2015improving}
Li HJ, Leung KS, Wong MH, Ballester PJ (2015) Improving {AutoDock Vina} using
  random forest: the growing accuracy of binding affinity prediction by the
  effective exploitation of larger data sets.
\newblock {\em Molecular informatics} 34(2-3):115--126.

\bibitem{wojcikowski2019development}
W{\'o}jcikowski M, Kukie{\l}ka M, Stepniewska-Dziubinska MM, Siedlecki P (2019)
  Development of a protein--ligand extended connectivity ({PLEC}) fingerprint
  and its application for binding affinity predictions.
\newblock {\em Bioinformatics} 35(8):1334--1341.

\bibitem{stepniewska2018development}
Stepniewska-Dziubinska MM, Zielenkiewicz P, Siedlecki P (2018) Development and
  evaluation of a deep learning model for protein--ligand binding affinity
  prediction.
\newblock {\em Bioinformatics} 34(21):3666--3674.

\bibitem{su2018comparative}
Su MY, et~al. (2018) Comparative assessment of scoring functions: The
  {CASF}-2016 update.
\newblock {\em Journal of chemical information and modeling} 59(2):895--913.

\bibitem{afifi2018improving}
Afifi K, Al-Sadek AF (2018) Improving classical scoring functions using random
  forest: The non-additivity of free energy terms' contributions in binding.
\newblock {\em Chemical biology \& drug design} 92(2):1429--1434.

\bibitem{PDBBind:2015}
Liu ZH, et~al. (2015) {{PDB}-wide collection of binding data: current status of
  the { PDBbind} database}.
\newblock {\em Bioinformatics} 31(3):405--412.

\bibitem{ollivier2007ricci}
Ollivier Y (2007) Ricci curvature of metric spaces.
\newblock {\em Comptes Rendus Mathematique} 345(11):643--646.

\bibitem{jost2014ollivier}
Jost J, Liu S (2014) Ollivier's {Ricci} curvature, local clustering and
  curvature-dimension inequalities on graphs.
\newblock {\em Discrete \& Computational Geometry} 51(2):300--322.

\bibitem{bhattacharya2015exact}
Bhattacharya BB, Mukherjee S (2015) Exact and asymptotic results on coarse
  {Ricci} curvature of graphs.
\newblock {\em Discrete Mathematics} 338(1):23--42.

\end{thebibliography}

\section*{Acknowledgments}

This work was supported in part by Nanyang Technological University Startup Grant M4081842.110, Singapore Ministry of Education Academic Research fund Tier 1 RG31/18 and RG109/19, and Tier 2 MOE2018-T2-1-033. 

\section*{Author contributions}
K.X. designed research; K.X. and J.J. performed research; K.X. and J.J. analyzed data; and K.X. and J.J. wrote the paper.

\section*{Supplementary materials}

\subsection*{Mathematical background of Ollivier Ricci curvature}

\subsubsection*{Lower bound of Ollivier-Ricci curvature}
Mathematically, Wasserstein distance is a linear minimization problem with convex constraints, hence it admits a dual formulation, i.e., Kantorovich duality. More specifically, let $W_1(m_x, m_y)$ be the Wasserstein distance between $m_x$ and $m_y$, i.e., two probability measurements as in Eq. (\ref{eq:probability}). Then $W_1(m_x, m_y)$ can be rewritten as,
	\begin{equation} \nonumber
	W_1(m_x, m_y) = \sup_{f\in Lip_1(\mathbb{R})}\bigg[\sum_{x_i\in V}f(x_i)m_x(x_i) - \sum_{x_i\in V}f(x_i)m_y(x_i) \bigg].
	\end{equation}
Here $f(x)$ is $1$-Lipschitz continuous $Lip_1(\mathbb{R})$, if it satisfies the following condition,
	\begin{equation} \nonumber
	|f(x)-f(y)| \le |x-y| = d(x,y) \quad \text{for all } x,y \in R.
	\end{equation}

For a locally finite graphs $G$, which may be infinite size but has finite degree for every vertex in $G$, its ORCs have lower bound  \cite{lin2010ricci,jost2014ollivier}. More specifically, for any pair of vertices $x$ and $y$ in $G$ and $x\sim y$, it satisfies that,
	\begin{equation}
	c(x,y) \ge 2\alpha -2 \max\bigg\{1-\frac{(1-\alpha)}{k_x} - \frac{(1-\alpha)}{k_y},0\bigg\}.
	\end{equation}
Note that here $d(x,y)=1$ if $x\sim y$. That means the graph has weight 1 for all its edges. From Kantorovich duality, one can have,
	\begin{align*} \nonumber
	&W_1(m_x, m_y)\\
	&= { \scriptstyle \sup_{f\in Lip_1(\mathbb{R})}\bigg[\sum \limits_{x_i\sim x}f(x_i)\frac{1-\alpha}{k_x} - \sum \limits_{x_i\sim y}f(x_i)\frac{1-\alpha}{k_y} +\alpha (f(x)-f(y))\bigg]} \\
	&= { \scriptstyle \sup_{f\in Lip_1(\mathbb{R})}\bigg[\sum \limits_{x_i\sim x, x_i\ne y}f(x_i)\frac{1-\alpha}{k_x} + f(y)\frac{1-\alpha}{k_x} - \sum\limits_{x_i\sim y, x_i\ne x}f(x_i)\frac{1-\alpha}{k_y} }\\
	&  { \scriptstyle  \qquad - f(x)\frac{1-\alpha}{k_y} +\alpha (f(x)-f(y))\bigg]}\\
	&= { \scriptstyle  \sup_{f\in Lip_1(\mathbb{R})}\bigg[\sum\limits_{x_i\sim x, x_i\ne y}(f(x_i)-f(x))\frac{1-\alpha}{k_x} - \sum\limits_{x_i\sim y, x_i\ne x}(f(x_i)-f(y))\frac{1-\alpha}{k_y} } \\
	&{ \scriptstyle  \qquad - \frac{k_y-1}{k_y}(1-\alpha)f(y)+\frac{k_x-1}{k_x}(1-\alpha)f(x) + f(y) \frac{1-\alpha}{k_x}}\\
	&{ \scriptstyle  \qquad -f(x)\frac{1-\alpha}{k_y} + \alpha(f(x)-f(y))\bigg]} \\
	&= { \scriptstyle  \sup_{f\in Lip_1(\mathbb{R})}\bigg[\sum\limits_{x_i\sim x, x_i\ne y}(f(x_i)-f(x))\frac{1-\alpha}{k_x} - \sum\limits_{x_i\sim y, x_i\ne x}(f(x_i)-f(y))\frac{1-\alpha}{k_y}}\\
	& { \scriptstyle  \qquad -\frac{1-\alpha}{k_x}(f(x)-f(y)) - \frac{1-\alpha}{k_y}(f(x)-f(y)) + f(x) - f(y)\bigg] }\\
	&{ \scriptstyle  \leq (1-\alpha)\frac{k_x-1}{k_x} + (1-\alpha)\frac{k_y-1}{k_y} +\bigg|1-\frac{(1-\alpha)}{k_x} - \frac{(1-\alpha)}{k_y}\bigg|}\\
	&= { \scriptstyle  (1-2\alpha)+ (1 - \frac{1-\alpha}{k_x} - \frac{1-\alpha}{k_y}) + \bigg|1-\frac{(1-\alpha)}{k_x} - \frac{(1-\alpha)}{k_y}\bigg|}\\
	&= { \scriptstyle 1-2\alpha + 2 \max\bigg\{1-\frac{(1-\alpha)}{k_x} - \frac{(1-\alpha)}{k_y},0\bigg\}}.
	\end{align*}
	Hence,
	\begin{align*} \nonumber
	 c(x,y) &= 1-W_1(m_x, m_y) \geq 2\alpha -2 \max\bigg\{{ \scriptstyle 1-\frac{(1-\alpha)}{k_x} - \frac{(1-\alpha)}{k_y},0} \bigg\} .
	\end{align*}

\subsubsection*{Ollivier-Ricci curvature for special graphs}

From the theorem above, several results can be derived, including upper bounds found for ORCs of some families of graphs  \cite{bhattacharya2015exact,jost2014ollivier,lin2010ricci}. In particularly, ORCs for four types special graphs, including tree graph, grid graph, complete graph and bipartite graph, are analyzed in great details.

\paragraph{ORC for tree graph}
A tree graph is a graph without any cycles. For an edge $e(x, y)$ from a tree graph, its edge ORC is,
\begin{equation}\label{tree}  \nonumber
c(x,y)=2\alpha -2 \max\bigg\{1-\frac{(1-\alpha)}{k_x} - \frac{(1-\alpha)}{k_y},0\bigg\}.
\end{equation}
To prove the above results, we can check its lower bound and upper bound. The lower bound for edge ORC can be directly attained from the above Kantorovich duality as,
	\begin{align*} \nonumber
	c(x,y) &\geq 2\alpha -2 \max\bigg\{1-\frac{1-\alpha}{k_x} - \frac{1-\alpha}{k_y},0\bigg\}.
	\end{align*}
For the upper bound of $c(x,y)$, we only need to consider the following two cases. i.e. $1-\frac{1-\alpha}{k_x} - \frac{1-\alpha}{k_y} < 0$ and $1-\frac{1-\alpha}{k_x} - \frac{1-\alpha}{k_y} \geq 0$.

If $1-\frac{1-\alpha}{k_x} - \frac{1-\alpha}{k_y} < 0$, at least one of the terms $\frac{1-\alpha}{k_x}$ and $\frac{1-\alpha}{k_y}$ is larger than 0.5. { Without the loss of generality}, we assume $\frac{1-\alpha}{k_x}>0.5$. According to Eq. (\ref{eq:probability}), that means all the weights for the adjacent vertices of node $x$ is larger than 0.5. Since the total weight is 1.0 and weights are nonnegative, the node $x$ can have only one adjacent vertex, i.e., vertex $y$ and $k_x=1$. We have probability measurements $m_x(y)=1-\alpha$ and $m_y(y)=\alpha$. If $1-\alpha>\alpha$, in any transportation plan, there is at least $(1-\alpha)-\alpha=1-2\alpha$ amount of ``mass" from $m_x$ that should be transported from vertex $y$ to its neighbour vertices (with distance 1), thus $W_1(m_x,m_y) \geq 1-2\alpha$. If $1-\alpha<\alpha$, that means $1-2\alpha<0$. It is obvious that $W_1(m_x,m_y) \geq 1-2\alpha$ as Wasserstein distance is nonnegative.

	Further, if $1-\frac{1-\alpha}{k_x} - \frac{1-\alpha}{k_y} \geq 0$, we can define an 1-Lipschitz continuous function $f(x_i)$ as follows \cite{jost2014ollivier},
	\begin{equation} \nonumber
	f(x_i) = \left\{ \begin{array}{ll}
	0, & x_i\sim y, x_i \ne x. \\
	1, & x_i = y. \\
	2, & x_i = x. \\
	3, & x_i \sim x, x_i \ne y.
	\end{array} \right.
	\end{equation}
	Then we have
	\begin{align*} \nonumber
	W_1(m_x, m_y) &\geq \frac{1-\alpha}{k_x}(3(k_x-1)+1) - \frac{2(1-\alpha)}{k_y}  +\alpha (2-1) \\
	&= 3 - 2\alpha - \frac{2(1-\alpha)}{k_x} - \frac{2(1-\alpha)}{k_y}\\
	&= 1-2\alpha + 2\bigg(1-\frac{(1-\alpha)}{k_x} - \frac{(1-\alpha)}{k_y}\bigg)
	\end{align*}
	Hence, combining both cases yields
	\begin{align*}
		W_1(m_x, m_y) \geq 1-2\alpha + 2\max\bigg\{1-\frac{(1-\alpha)}{k_x} - \frac{(1-\alpha)}{k_y},0\bigg\}.
	\end{align*}
	Therefore,
	\begin{align*}
	c(x,y) &=  2\alpha - 2\max\bigg\{1-\frac{(1-\alpha)}{k_x} - \frac{(1-\alpha)}{k_y},0\bigg\}.
	\end{align*}
The above theorem has been proven for the special case of $\alpha=0.0$  \cite{jost2014ollivier}. In our computation, we use $\alpha=0.5$, thus we have $c(x,y)=\frac{1}{k_x} + \frac{1}{k_y} - 1$. Note that if $k_x$ and $k_y$ approaches $\infty$, $c(x,y)$ approaches $-1$.

\paragraph{ORC for grid graph}
For an infinite-sized grid graph, all ORCs are zero. For a finite-sized grid graph, other than edges with vertex (either one or two) from boundary points, all the other edges have zero ORCs. This is due to the fact that the cost of moving $m_x$ to $m_y$ is equal to $d(x,y)$. Hence,
	\begin{equation} \nonumber
	c(x,y) = 1 - \frac{W_1(m_x, m_y)}{d(x,y)} = 1 - \frac{d(x,y)}{d(x,y)} = 0.
	\end{equation}

\paragraph{ORC for complete graph}

For a graph $G$, if $\sharp(x,y)$ represents the number of triangles with $x$ and $y$ being vertices (of the triangles) in $G$, edge ORC for $e(x,y)$ satisfies
	\begin{multline} \nonumber
	{ \scriptstyle  c(x,y) \geq -\max\bigg\{1-\frac{1-\alpha}{k_x}-\frac{1-\alpha}{k_y}-(1-\alpha)\frac{\sharp(x,y)}{\min(k_x, k_y)},0\bigg\} }\\
	{ \scriptstyle   -\max\bigg\{1-\frac{1-\alpha}{k_x}-\frac{1-\alpha}{k_y}-(1-\alpha)\frac{\sharp(x,y)}{\max(k_x, k_y)},0\bigg\}}
	{ \scriptstyle  +(1-\alpha)\frac{\sharp(x,y)}{\min(k_x,k_y)} + 2\alpha.}
	\end{multline}
Again, the inequality above was proven for $\alpha=0.0$ in  \cite{jost2014ollivier} and its proof can be modified similarly for all values of $\alpha$. The inequality is also sharp for complete graphs. That is,
	\begin{equation} \nonumber
	c(x,y) = (1-\alpha)\frac{n}{n-1}.
	\end{equation}

For a complete graph with $n\ge 3$, we have $\sharp(x,y) = n-2$ and $k_x=k_y=n-1$ for any $x$ and $y$ in $G$. Hence,
	\begin{align*} \nonumber
	c(x,y)
	&={ \scriptstyle -2\bigg(1-\frac{1-\alpha}{n-1}-\frac{1-\alpha}{n-1}-(1-\alpha)\frac{n-2}{n-1}\bigg) + (1-\alpha)\frac{n-2}{n-1} + 2\alpha}\\
	&= -2 +\frac{2n(1-\alpha)}{n-1} +\frac{n-2}{n-1}(1-\alpha)  +2\alpha\\
	&= -2 + (1-\alpha)\frac{n}{n-1} + 2(1-\alpha) + 2\alpha\\
	&= (1-\alpha)\frac{n}{n-1}.
	\end{align*}
In our computation with $\alpha=0.5$, we have $c(x,y)=\frac{n}{2(n-1)}$.

\paragraph{ORC for bipartite graph}
Recently, a closed formula is found for the ORC of weighted bipartite graphs  \cite{bhattacharya2015exact}. { Similarly, the closed formula can be adapted to our probability measure $m_x$, which provides a special case in terms of $\alpha$.} Let $G=(V,E)$ be a locally finite weighted bipartite graph and $e(x,y) \in E$. Suppose $R(x,y)$ is a subgraph of $G_{(x,y)}$ induced by $N_1(x)\cup N_1(y)$, and $R_1(x,y)$, $R_2(x,y)$, $\cdots$, $R_q(x,y)$ be the connected components of $R(x,y)$. If $U_a(x)=V(R_a(x,y)) \cap N_1(x)$ and $U_a(y) = V(R_a(x,y))\cap N_1(y)$ for $a\in\{1,2,\cdots, q\}$, then
	\begin{equation} \nonumber
	\begin{split}
	c(x,y) &= 2\alpha -2\max\bigg\{1-\frac{(1-\alpha)}{k_x} -\frac{(1-\alpha)}{k_y}- \frac{|N_1(y)|(1-\alpha)}{k_y}
	\\ &\qquad\qquad +\sum_{a=1}^{q} \max\bigg\{{ \scriptstyle \frac{|U_a(y)|(1-\alpha)}{k_y} - \frac{|U_a(x)|(1-\alpha)}{k_x},0}\bigg\},0\bigg\}.
	\end{split}
	\end{equation}
	\label{Bipartite_formula}
Note that the above closed formula is symmetric, i.e. $c(x,y) = c(y,x)$. Edge ORC for any $e(x,y) \in E(G)$ depends only on the induced subgraph $G_{(x,y)}$   \cite{bhattacharya2015exact}.

For a complete bipartite graph $K_{m,n}$, in which two vertex sets are composed of $m$ and $n$ vertices respectively, and edges exist between any two vertices from different sets, we have the following simplified closed formula,
\begin{align*} \nonumber
\begin{split}
 c(x,y)
&=2\alpha -2\max\{1-\frac{(1-\alpha)}{m} -\frac{(1-\alpha)}{n}- \frac{(m-1)(1-\alpha)}{m}
\\ &\qquad\qquad +\max\{\frac{(m-1)(1-\alpha)}{m} - \frac{(n-1)(1-\alpha)}{n},0\},0\}
\end{split}
\\
 &= { \scriptstyle  2\alpha -2\max\{ 1-\frac{(1-\alpha)}{m} -\frac{(1-\alpha)}{n} -\min\{ \frac{(m-1)(1-\alpha)}{m}, \frac{(n-1)(1-\alpha)}{n} \},0 \}}.
\end{align*}
Note that when $\alpha=0.5$, we have $c(x,y)=\min\{\frac{1}{m}, \frac{1}{n}\}$. Moreover, if we have $m=n$, then ORCs satisfy,
\begin{align*} \nonumber
c(x,y)
&= 2\alpha -2\max\bigg\{ { \scriptstyle 1-\frac{(1-\alpha)}{n} -\frac{(1-\alpha)}{n} -\frac{(n-1)(1-\alpha)}{n},0 }\bigg\}\\
&= 2\alpha -2\max\bigg\{ \alpha- \frac{(1-\alpha)}{n},0\bigg\}.
\end{align*}
We have $c(x,y)=\frac{1}{n}$ if $\alpha=0.5$.

\subsection*{Ollivier-Ricci curvature for osmolyte hydrogen-bonding network analysis}

\paragraph{Molecular dynamic simulation of osmolyte systems}
Using GROMACS-5.1.2, molecular dynamic simulations were performed on TMAO Kast Model and urea (Model from AMBER package) with the four point (TIP4P-EW) water model. Pure water with concentrations of urea and TMAO from 1M to 8M are simulated in the water model for 100 nanoseconds (ns). 3000 water molecules are maintained in all cases. Using the insert-molecules utility in GROMACs, we randomly place the urea/TMAO molecules in the configurations and apply random insertion of 3000 water molecules in a simulation cube. Hence, for each concentration, the water molecules remain the same but the urea/TMAO molecules vary. Under NVT conditions, an equilibration is performed with temperature 300K for 10 picoseconds (ps) followed by another 100 ps under NPT conditions but using 2 frame per second (fs) timestep, Berendsen thermostats with $\tau=0.1$ ps and barostat $\tau=2$ ps. Constraining of bonds and its angles by LINCS algorithm is used. In order to achieve the specified concentrations within the simulation time of 2 ps during the NPT phase and without constraints, calibration is performed between volume of cube and the amount of urea/TMAO molecules. The production run is then repeated three times for 100 ns with the Berenden thermostat (temperature 300K and $\tau =0.1$ ps) and Parrinello-Rahman barostat (pessure = 1 bar and $\tau = 2$ ps) with time step of 2 fs. Furthermore, the Newton's equation of motion is integrated using a leap-frog algorithm. Note that the cut-offs for the van der Waals (vdW) interactions and the short-range electrostatic interactions are fixed at 1.0 nanometre (nm) with a particle mesh Eswald (PME) method applied to the long-range electrostatic interactions.

\paragraph{ORC for hydrogen-bonding network characterization}

A total 101 configurations are extracted equally from the molecular dynamic (MD) simulations for each concentration of TMAO and urea. For each configuration, a hydrogen bonding network is generated by using all the O atoms from water molecules and a cutoff distance 4.0\AA. Vertex and edge ORCs can be calculated. Other than their statistic properties, we can also study their distributions. For a better characterization and comparison, we consider a density function (DF) $\frac{1}{nh} \sum_{i=1}^n K(\frac{c-c_i}{h})$ for ORC series $\{c_i;i=1,2,...,n\}$. Here $K$ is a non-negative kernel function with scale parameter $h>0$. In our computation, we use $K(\frac{c-c_i}{h}) = \frac{1}{\sqrt{2\pi}}e^{-\frac{(c-c_i)^2}{2h^2}}$ and $h=1.059\min\{\sigma_C,\text{IQR}/1.34\}n^{-\frac{1}{5}}$ with $\sigma_C$ the standard deviation of $C$ and IQR the Interquartile Range of $C$. Using this density function, we can systematically compare the behaviors of the two osmolyte systems at different concentrations.

The comparison of density distributions of two osmolyte systems, obtained from the last frame of MD simulations for all the eight concentrations, are demonstrated in Figure \ref{fig:last-frame}. Similar to average results in Figure \ref{fig:osmolyte}, as ion concentration increases, the peaks of urea DFs have only slight decrease, while peaks of TMAO DFs undergo dramatic decrease. This is most obvious in vertex ORC based DFs.

In addition, Figures \ref{fig:TMAO-3d} and \ref{fig:urea-3d} shows the 3D representation of a total 101 vertex-ORC-based DFs for all eight concentrations of TMAO and urea respectively. Again, we can see that, as ion concentration increases, there is less variation of DF peaks for urea systems, while the peaks of TMAO have more variation across all 101 frames. In both figures, we generally see a change in red regions as critical peaks decreases from red to green.

\renewcommand{\thefigure}{S\arabic{figure}}
\setcounter{figure}{0}

\begin{figure}[ht]
	\centering
	\includegraphics[width=0.9\linewidth]{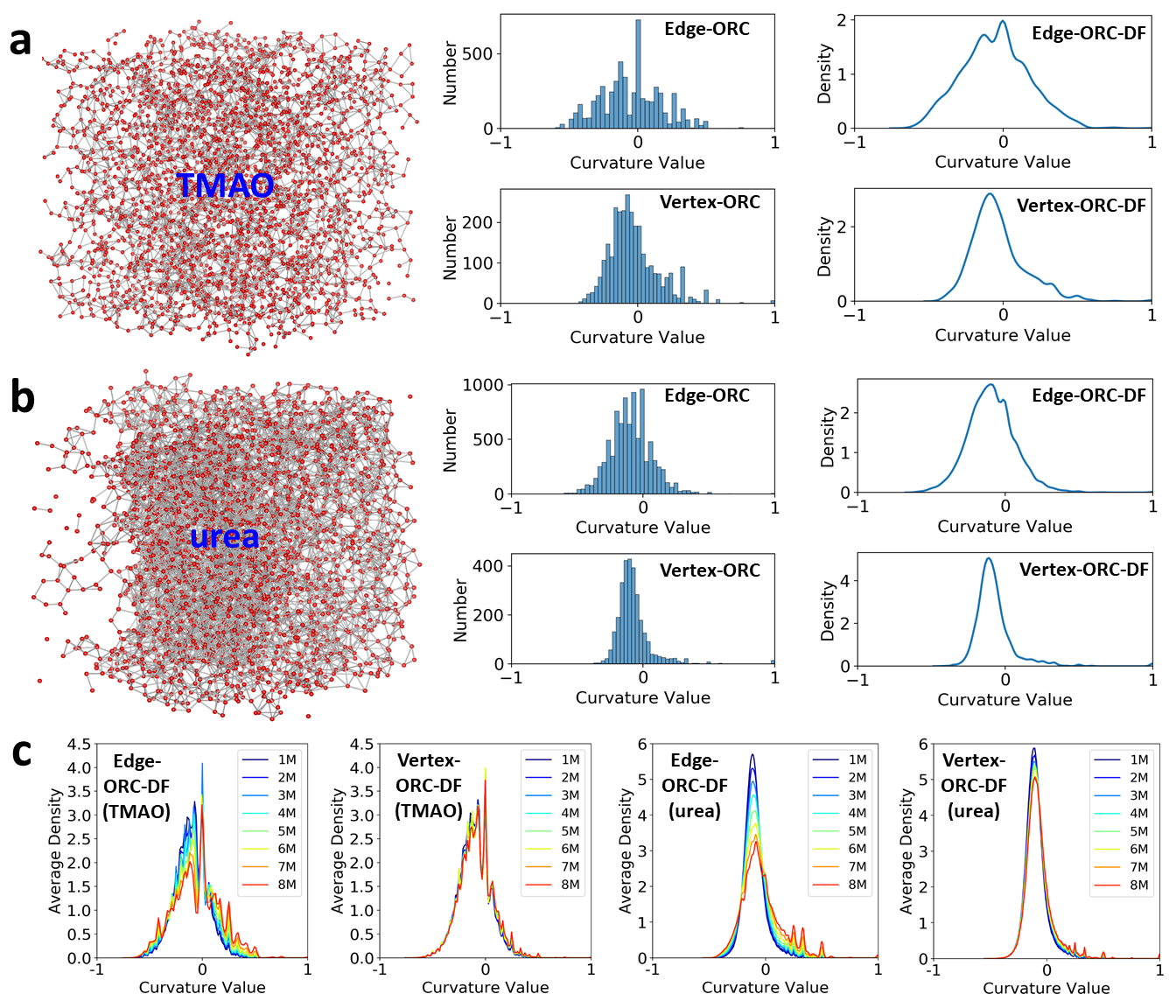}
	\caption{Illustration of ORC-based quantitative characterization of hydrogen bonding network in TMAO and urea. {\bf a} TMAO graph is generated from last frame of its MD simulation (ion concentration 8M), with a cutoff distance 4\AA. The distribution of edge and vertex  ORCs are transferred into the corresponding distribution functions (DFs) using kernel density estimator. {\bf b} The urea graph is generated from last frame of its MD simulation (ion concentration 8M), with a cutoff distance 4\AA. Their edge and vertex ORC-DFs are generated in the same way as TMAO. {\bf c} The comparison of the average TMAO and urea DFs in 8 different concentrations (1M to 8M). It can be seen that there is a significant difference in the variation of edge and vertex ORCs between TMAO and urea across different ion concentrations.}
	\label{fig:osmolyte}
\end{figure}

\begin{figure} [ht]
	\centering
	\includegraphics[width=.8\linewidth]{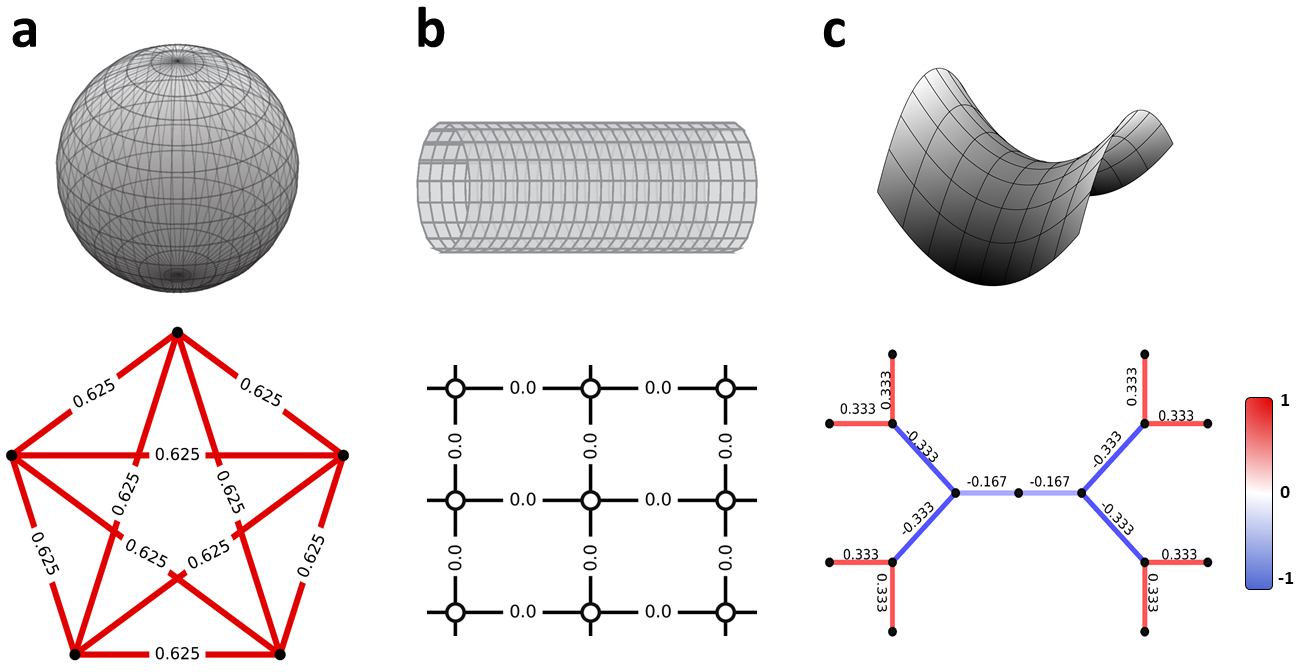}
	\caption{Illustration of Ollivier-Ricci curvatures for three graph representations from three basic types of quadratic surfaces, i.e., elliptic ({\bf a}), parabolic ({\bf b}), and hyperbolic ({\bf c}). {\bf a} Graph from parabolic surface has only positive ORCs. {\bf b} Infinitely-sized grid graph has only zero ORCs. {\bf c} Negative ORCs are found on link or bridge regions.}
\label{fig:three_types}
\end{figure}

\begin{figure}[ht]
	\centering
	\includegraphics[width=0.9\textwidth]{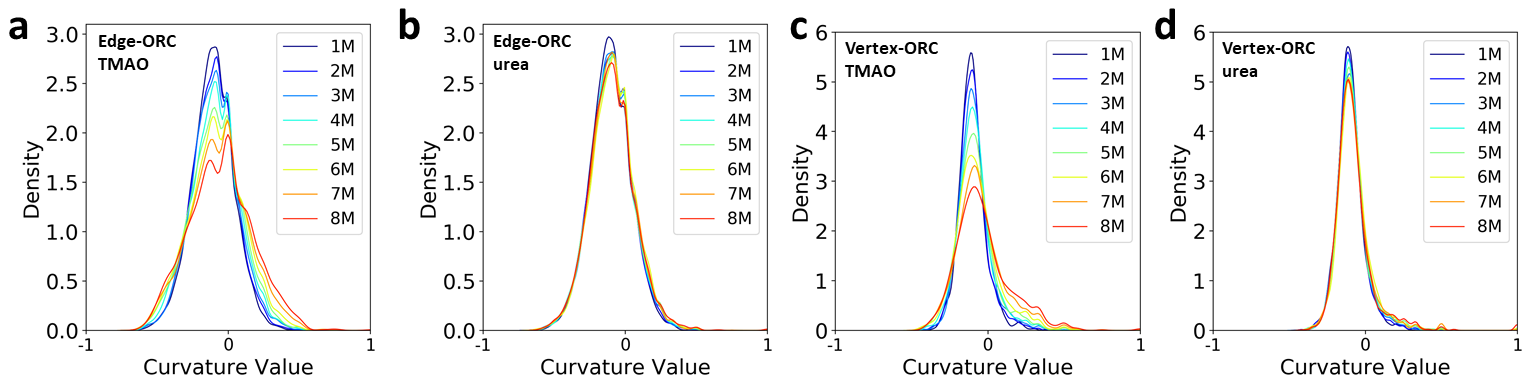}
	\caption{Comparisons of density distributions of edge and vertex ORCs between TMAO and urea from the last configuration of MD simulation from concentration 1M to 8M. {\bf a} Density functions of edge ORCs of urea. {\bf b} Density functions of edge ORCs of TMAO. {\bf c} Density functions of vertex ORCs of urea. {\bf d} Density functions of vertex ORCs of TMAO.}
	\label{fig:last-frame}
\end{figure}

\begin{figure}[ht]
	\centering
	\includegraphics[width=0.8\textwidth]{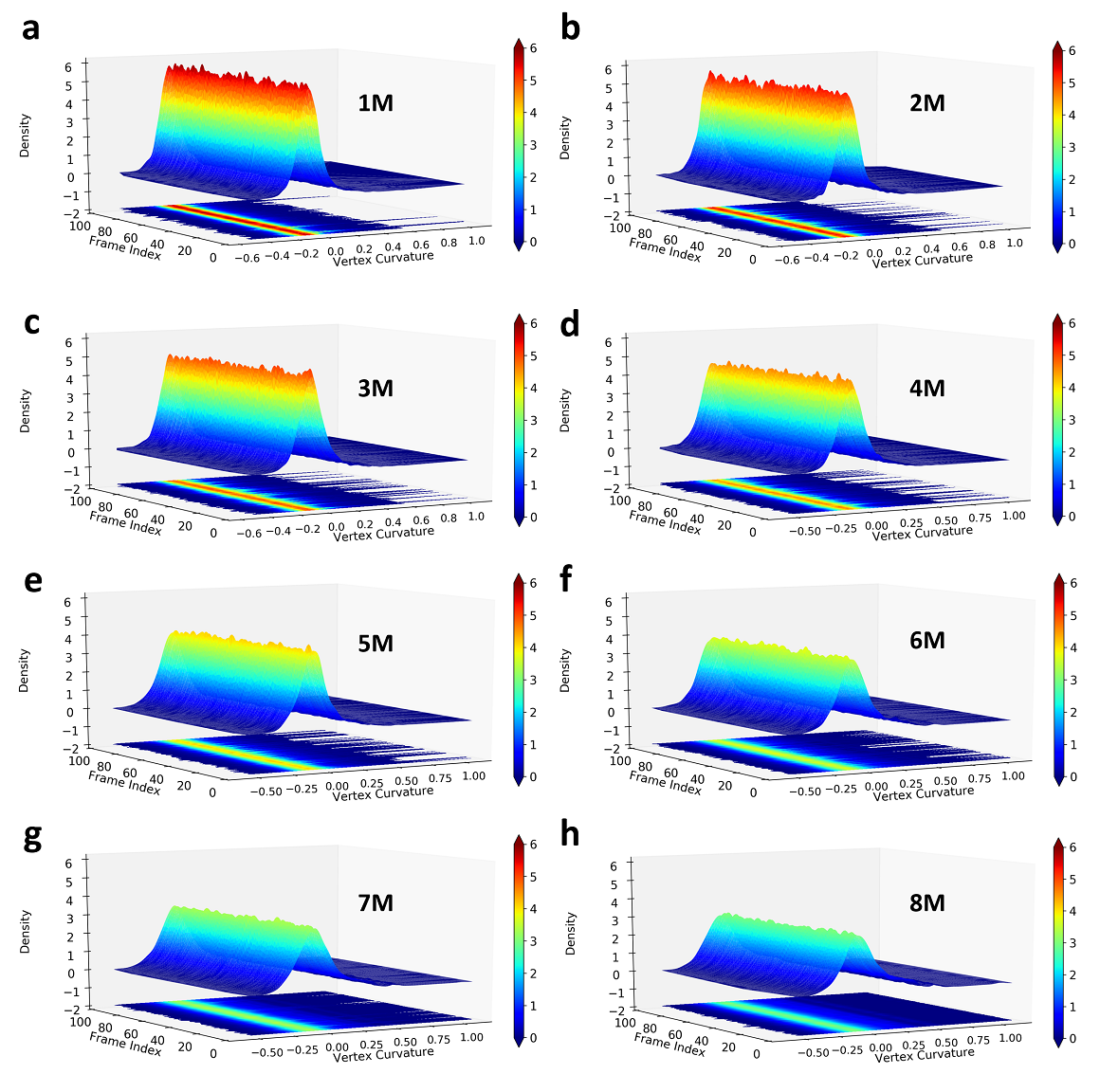}
	\caption{Vertex-ORC based TMAO density functions for all 101 frames from the MD simulations. {\bf a}-{\bf h} corresponds to 1M to 8M ion concentrations.}
	\label{fig:TMAO-3d}
\end{figure}

\begin{figure}[ht]
	\centering
	\includegraphics[width=0.8\textwidth]{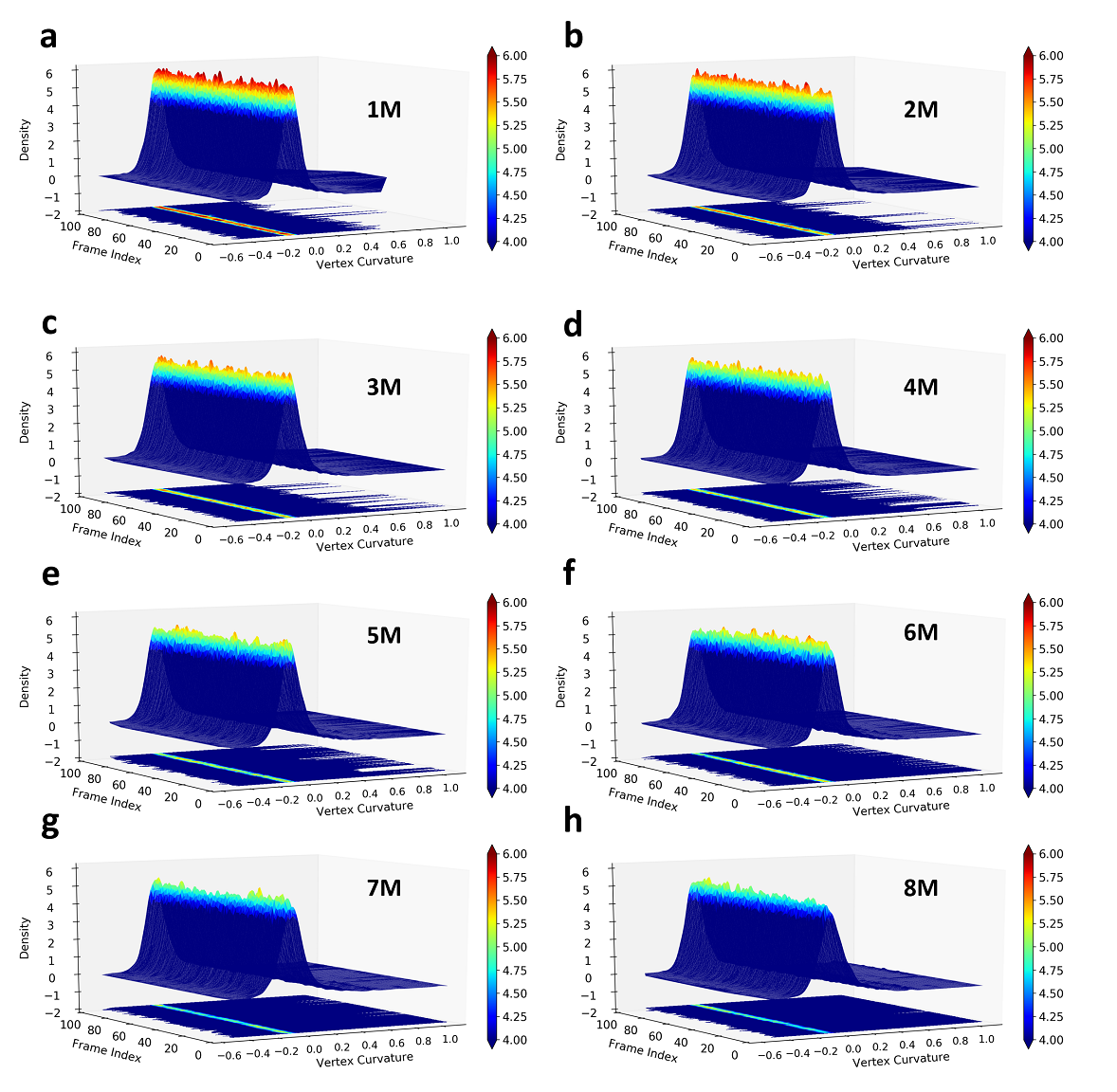}
	\caption{Vertex-ORC based urea density functions for all 101 frames from the MD simulations. {\bf a}-{\bf h} corresponds to 1M to 8M ion concentrations.}
	\label{fig:urea-3d}
\end{figure}

\renewcommand{\thetable}{S\arabic{table}}
\setcounter{table}{0}

\begin{table}[ht]
	\centering
	\caption{Detailed information of PDBbind-v2007, PDBbind-v2013 and PDBbind-v2016 Databases.}
	\begin{tabular}{|c|c|c|c|}
		\hline
		\textbf{Version} & \textbf{Refined set} & \textbf{Training set} & \textbf{Core set (Test set)} \\ \hline
		PDBbind-v2007            & 1300                 & 1105                  & 195                          \\ \hline
		PDBbind-v2013            & 2959                 & 2764                  & 195                          \\ \hline
		PDBbind-v2016            & 4057                 & 3772                  & 285                          \\ \hline
	\end{tabular}

	\label{tab:PDBbind}
\end{table}

\begin{table}[ht]
	\centering
	\caption{The parameters in our GBT model. Note that 10 repetitions are considered in our model.}
	\begin{tabular}{cccc}
		\hline
		\multicolumn{1}{|c|}{\textbf{No. of Estimators}} & \multicolumn{1}{c|}{\textbf{Max Depth}}    & \multicolumn{1}{c|}{\textbf{Min. Sample Split}} & \multicolumn{1}{c|}{\textbf{Learning Rate}} \\ \hline
		\multicolumn{1}{|c|}{40000}                      & \multicolumn{1}{c|}{7}                     & \multicolumn{1}{c|}{2}                          & \multicolumn{1}{c|}{0.001}                  \\ \hline
		\multicolumn{1}{|c|}{\textbf{Loss Function}}     & \multicolumn{1}{c|}{\textbf{Max Features}} & \multicolumn{1}{c|}{\textbf{Subsample Size}}    & \multicolumn{1}{c|}{\textbf{Repetition}}    \\ \hline
		\multicolumn{1}{|c|}{Least Squares}              & \multicolumn{1}{c|}{Square Root}           & \multicolumn{1}{c|}{0.7}                        & \multicolumn{1}{c|}{10 times}               \\ \hline
	\end{tabular}

	\label{tab:GBT-hyperparam}
\end{table}

\end{document}